\documentclass[aps,pra,twocolumn,showpacs,10pt,floatfix]{revtex4-1}

\usepackage{graphicx}
\usepackage{dcolumn}
\usepackage[english]{babel}
\usepackage{amsmath}
\usepackage{amssymb}
\usepackage{tabularx}
\usepackage{multirow}
\usepackage[all]{xy}
\usepackage{appendix}
\usepackage{xspace}
\usepackage{color}
\usepackage{array}

\begin{document}
\title{Revisiting interferences for measuring and optimizing optical nonlinearities}

\author{F. Billard}
\author{P. B\'ejot}
\author{E. Hertz}
\author{B. Lavorel}
\author{O. Faucher}

\affiliation{Laboratoire Interdisciplinaire CARNOT de Bourgogne, UMR 6303 CNRS-Universit\'e de Bourgogne, BP 47870, 21078 Dijon, France}

\begin{abstract}
A method based on optical interferences for measuring optical nonlinearities is presented. In a proof-of-principle experiment, the technique is applied to the experimental determination of the intensity dependence of the photoionization process. It is shown that it can also be used to control and optimize the nonlinear process itself at constant input energy. The presented strategy leads to enhancements that can reach several orders of magnitude for highly nonlinear processes.
\end{abstract}

 \maketitle
\section{Introduction}
Since the discovery of the laser, technological progresses have allowed to produce optical sources of constantly growing power. Laser intensities can now be so high that atoms and molecules, when exposed to such electric fields, can absorb simultaneously a large number of photons, leading to highly nonlinear dynamics and to the observation of phenomena such as strong field ionization or high-order harmonic \cite{Ferray88} and attosecond pulse \cite{Antoine96} generation. More particularly, at moderate intensity, the efficiency of these processes scales as a fixed power of the intensity. For instance, the rate of  multiphoton ionization is well approximated by $I^N$, where $I$ is the laser intensity and $N$ is the minimal number of photons that the electronic wavepacket needs to absorb for escaping from the ionic attractive potential. As the intensity rises (typically higher than a few 10 TW/cm$^2$), the electronic structure of the system is strongly distorted allowing the occurrence of processes such as tunnel ionization, above-threshold ionization, and ionization channel closure. As a consequence, the very simple relation linking laser intensity and ionization does not hold anymore. Instead, only an effective nonlinearity, which can also depend on the intensity, can be extracted from experiments. For instance, an effective nonlinearity $N_{\textrm{eff}}\simeq$7.5 \cite{KasparianChin00,LoriotIonization} has been determined for the  ionization rate of argon at 800 nm for $I\simeq$ 50 TW/cm$^2$, while a nonlinearity of 11 is expected in the multiphoton regime. The nonlinearity of a process $\mathcal{P}$ is generally determined by fitting the measured process yield by a function proportional to $I^{N_{\textrm{eff}}}$. Unfortunately, due to the highly nonlinear nature of the process, the dynamic range needed for an accurate measurement of such an effective nonlinearity is hardly achievable experimentally and leads to large uncertainties in the determination of $N_{\textrm{eff}}$. In this paper, a simple method to determine the nonlinearity of any optical processes is implemented. This method is based on optical interferences induced by two crossed laser beams and relies only on two measurements in contrast with standard methods which need to reconstruct the dependence of the process as a function of the intensity on several decades. The presented method, which can be applied to any nonlinear process, is demonstrated analytically, numerically, and experimentally in the case of ionization. The proposed strategy also provides a control and enhance the production efficiency of the nonlinear process. In particular, we show that the yield can be increased by several order of magnitude and that the overall energy needed for the production of any nonlinear process in the two-beam geometry is reduced up to a factor 2 as compared to the single beam geometry. In the first section, the principle of the method is presented analytically and compared to full 3D+1 pulse nonlinear propagation simulations. In the second section, experimental results are presented and compared to theoretical results in several gases, leading to the determination of the effective nonlinearity driving the ionization process in these gases.

\section{Theoretical approach}

The method presented in this paper is based on the fact that any nonlinear process is enhanced when, at constant energy level, two coherent pulses spatially interfere as compared to a single pulse experiment. In the first subsection, the principle is demonstrated with infinite plane waves. In the second subsection, the concept is extended analytically to gaussian beams propagating linearly. Finally, the result is compared to realistic full 3D+1 nonlinear propagation simulations. The comparison between analytical calculations and numerical simulations is of particular importance when dealing with bulk media since nonlinear effects usually lead to strong deviations between analytical results obtained in the linear regime and realistic propagation resulting in large uncertainty on the experimental determination of the nonlinearity.

\subsection{Plane waves case}
\begin{figure}[htbp!]
\begin{center}
      \includegraphics[keepaspectratio,width=8cm]{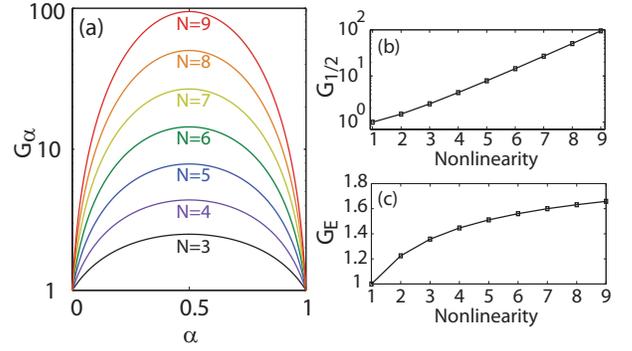}
\end{center}
   \caption{(Color online) (a) Gain on the nonlinear process as a function of $\alpha$ for different nonlinearity orders $N$. Optimal gain on the nonlinear process (b) and energy gain (c) as a function of the nonlinearity.}
\label{Fig:1}
\end{figure}

In this section, analytical gain on a nonlinear process is derived when two crossing plane waves with same carrier frequencies interfere. Assuming infinitely extended plane waves, the electric field envelope $\varepsilon$ composed of two distinct crossing fields writes

\begin{equation}
\varepsilon=\varepsilon_1+\varepsilon_2e^{i\phi},
\end{equation}
where $\varepsilon_1$ (resp. $\varepsilon_2$) is the electric field envelope of the first (resp. second) pulse and $\phi$ is the relative phase between the two pulses. If one defines $\varepsilon_1$ (resp. $\varepsilon_2$) such that $|\varepsilon_1|^2=I_1$ (resp. $|\varepsilon_1|^2=I_1$) where $I_1$ (resp. $I_2$) is the intensity of the first (resp. second) beam, the total intensity then writes

\begin{eqnarray}
I&=&I_1+I_2+2\varepsilon_1\varepsilon_2\cos\phi\\ \nonumber
 &=&I_\textrm{T}\left(1+2\sqrt{\alpha(1-\alpha)}\cos\phi\right),
\end{eqnarray}
with $I_\textrm{T}=I_1+I_2$ the total intensity assuming no interference and $\alpha=I_1/I_\textrm{T}$ the relative intensity of the first pulse as compared to $I_\textrm{T}$. Assuming that the studied nonlinear process $\mathcal{P}$ is directly proportional to the $N^{\textrm{th}}$ power of intensity, the averaged signal over an interference fringe writes
\begin{equation}
S_\alpha\propto \frac{I^N_\textrm{T}}{2\pi}\int_0^{2\pi}\left(1+C_\alpha\cos\phi\right)^Nd\phi,
\end{equation}
with $C_\alpha=2\sqrt{\alpha(1-\alpha)}$ the fringe contrast. The signal $S_1$ induced by a single beam with the same averaged intensity is given by

\begin{equation}
S_1\propto I^N_\textrm{T}.
\end{equation}
The gain $G_\alpha$ then writes
\begin{equation}
G_\alpha=\frac{S_\alpha}{S_1}=\ ^2\!F_1\left(\frac{1-N}{2}, -\frac{N}{2}, 1, C_\alpha^2\right),
\end{equation}
where $^2\!F_1$ is the hypergeometric function.
Figure \ref{Fig:1}(a) shows the gain $G_\alpha$ as a function of $\alpha$ for several nonlinearities. The gain on the nonlinear process can easily be controlled by adjusting the relative energy of the two pump beams. The maximal gain is obtained when the energy is equally shared between the two beams
($\alpha=0.5)$, \textit{i.e.} when the fringe contrast is maximal ($C=1$). In this case, the gain $G_{1/2}$ compared to a single pulse of intensity $I_\textrm{T}$ writes
\begin{equation}
G_{1/2}=\frac{\Gamma\left(2N+1\right)}{2^N\Gamma^2\left(N+1\right)},
\label{Eq:Gplane}
\end{equation}
where $\Gamma: z\mapsto \int_0^{+\infty}  t^{z-1}\,e^{-t}\,\mathrm{d}t$ denotes the gamma function. As shown in Fig. \ref{Fig:1}(b), the obtained gain is monotonic with the nonlinearity $N$ so that the determination of the gain leads in turn to the unambiguous determination of the nonlinearity. Using this method, the determination of the nonlinearity only requires two measurements ($S_1$ and $S_{1/2}$) unlike conventional methods which need to reconstruct the full dependence of the nonlinear process as a function of the intensity.

Note that an equivalent nonlinear signal would be generated with a single pulse of intensity $I_\textrm{eq}$=$G_\textrm{E}I_\textrm{T}$ where $G_\textrm{E}$ is the energy gain. One can show that

\begin{equation}
G_\textrm{E}=\left(G_{1/2}\right)^{1/N}\underset{\substack{N \to \infty}}\sim 2-\frac{\ln(\pi N)}{N}\underset{\substack{N \to \infty}}{\rightarrow} 2
\end{equation}

As a consequence, the overall energy needed for the production of a nonlinear process in the two-beam geometry can be reduced up to a factor 2 as compared to the single beam geometry.

\subsection{Gaussian waves case}

In the previous section, the gain on the generation of any nonlinear process has been derived for infinite plane waves. Considering plane waves is a useful approximation for understanding the proposed method and provides very simple analytical formula. Nevertheless, most of nonlinear processes are produced with non collimated beams. In this section, the gain on a nonlinear process is therefore derived in the case of gaussian beams undergoing a linear propagation. The results are then compared to the gain numerically calculated with a 3D+1 nonlinear propagation equation.

\subsubsection{Analytical considerations}
In the case of two gaussian beams of wavenumber $k_0$ crossing with an angle $2\theta$, the intensity profile in the bisecting plane ($x$,$z$) as a function of the propagation distance $z$ is given by
\begin{equation}
I(x,z)=I_1(x,z)+I_2(x,z)+I_{\textrm{cross}}(x,z)\cos{\left(2k_0\sin{\theta}x\right)},
\end{equation}
with

\begin{eqnarray}
I_1(x,z)&=&\frac{\alpha I_\textrm{T}}{|A|^2(z)}\exp{\left(\frac{-2(x-x_0(z))}{\sigma^2_x(z)}^2\right)}\\ \nonumber
I_2(x,z)&=&\frac{(1-\alpha)I_\textrm{T}}{|A|^2(z)}\exp{\left(\frac{-2(x+x_0(z))}{\sigma^2_x(z)}^2\right)}\\ \nonumber
I_{\textrm{cross}}(x,z)&=&\frac{2\sqrt{\alpha(1-\alpha)}I_\textrm{T}}{|A|^2(z)}\exp{\left(\frac{-2(x^2+x^2_0(z))}{\sigma^2_x(z)}\right)}
\end{eqnarray}

and

\begin{eqnarray}
A(z)&=&\frac{1}{1+z/q_z} \\ \nonumber
x_0(z)&=& z\tan{\theta}\\ \nonumber
1/q_z&=& 1/R_z-i\lambda_0/(\pi\sigma^2_0) \\ \nonumber
\sigma_x(z)&=& \sigma_0\sqrt{1+(z/z_r)^2} \\ \nonumber
R_z&=&z(1+(z_r/z)^2) \\ \nonumber
z_r&=&\pi\sigma^2_0/\lambda_0,
\end{eqnarray}
where $\lambda_0$ is the wavelength, $q_z$ is the complex radius, $\sigma_0$ is the beam waist, $I_\textrm{T}$ is the total intensity, and $z_r$ is the Rayleigh length.
In the particular case  $\alpha$=0.5, it reduces to

\begin{equation}
I=2\frac{I_\textrm{T}}{|A|^2(z)}\exp{\left(-\frac{2(x^2+x^2_0)}{\sigma^2_x(z)}\right)}\left|\cosh{\left(Bx\right)}\right|^2,
\end{equation}
with $B=\frac{2x_0(z)}{\sigma^2_x(z)}-ik_0\sin\theta$.
\\

The gain on the nonlinear process integrated over the propagation distance is given by

\begin{equation}
G_\alpha=\frac{\iint{I^N(x,z)dxdz}}{\frac{1}{\alpha^N}\iint{I_1^N(x,z)dxdz}}.
\label{Eq:GaussianWave}
\end{equation}

\begin{figure}[htbp!]
\begin{center}
      \includegraphics[keepaspectratio,width=8cm]{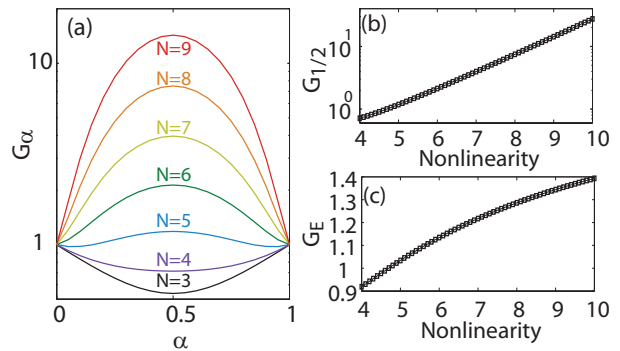}
\end{center}
   \caption{(Color online) (a) Gain on the nonlinear process as a function of $\alpha$ for several nonlinearity orders in the case of gaussian beams. Optimal gain on the nonlinear process (b) and energy gain (c) as a function of the nonlinearity.}
\label{Fig:2}
\end{figure}

Figure \ref{Fig:2}(a) shows the gain $G_\alpha$ as a function of $\alpha$ for several nonlinearities. The initial conditions were chosen to be $\lambda_0$=800 nm, $\sigma_0$=22 $\mu $m, and $\theta$=3$^\circ$. The gain integrated over the propagation length $z$ is considerably reduced as compared with the plane wave analytical solutions. This reduction is diminished when the interaction region is limited by the length of the nonlinear medium itself rather than by the crossing effect. For low nonlinearities, the generation of the nonlinear process is even reduced as compared with the single beam case. Figures \ref{Fig:2}(b,c) display $G_{1/2}$ and the energy gain $G_\textrm{E}$ as a function of the nonlinearity order calculated with optimal initial conditions, \textit{i.e.} $\alpha=0.5$. As it is the case for plane waves, one can show that $G_E\underset{\substack{N \to \infty}}{\rightarrow} 2$. As a consequence, the same amount of nonlinear signal can be obtained in the two-pulse scheme with a total energy down to half the one required in the single pulse geometry. Moreover, the presented method leads to very efficient optimization (about one order of magnitude for nonlinearity $N$=8-9) of the nonlinear process generation, in particular if one compares with standard control strategy based on optical pulse shaping \cite{Brixner}.

\subsubsection{Numerical considerations}
We chose to test the presented method with the photoionization process. In this case, the interferences between the two pulses lead to the formation of a plasma grating. Such a grating has been the subject of rising interest. For instance, it has been shown that a moving plasma grating formed by two intersecting lasers of slightly different central frequency leads to an energy exchange between them \cite{GratingEnergyExchange} or that a stationary grating formed with pulses of same frequency results in the divergence or in the redirection of a third laser beam \cite{GratingDiffraction}. In order to assess the analytical results derived above and evaluate the influence of nonlinear propagation effects such as intensity clamping in realistic experimental conditions, full 3D+1 numerical propagation calculations were performed. The propagation in both single ($\alpha=1$, $E=50~\mu$J) and two-beam ($\alpha=0.5$, $E_1=E_2=25~\mu $J) geometries were considered, allowing the comparison of the amount of electrons generated all along the propagation. The equation driving the propagation of the global electric field envelope $\varepsilon$ reads \cite{UPPE_Moloney}

\begin{equation}
\partial_z\widetilde{\varepsilon}=i\left(k_z-k_1\omega\right)\widetilde{\varepsilon}+\frac{1}{k_z}\left(\frac{i\omega^2}{c^2}\widetilde{P}_\textrm{NL}-\frac{\omega}{2\epsilon_0c^2}\widetilde{J}\right)-\widetilde{L}_\textrm{losses},
\label{Eq:final}
\end{equation}
with
\begin{eqnarray}
k(\omega)&=&n(\omega)\omega/c,  k_z=\sqrt{k^2(\omega)-(k_x^2+k_y^2)},\\ \nonumber
P_\textrm{NL}&=&n_{2}|\varepsilon|^{2}\varepsilon,  \partial_t\rho=\sigma_NI^N(\rho_\textrm{at}-\rho),\\ \nonumber
\widetilde{J}&=&\frac{e^2}{m_\textrm{e}}\frac{\nu_\textrm{e}+i\omega}{\nu_\textrm{e}^2+\omega^2}\widetilde{\rho\varepsilon},\\ L_\textrm{losses}&=&\frac{N\hbar\omega_0\sigma_N\rho_\textrm{at}}{2}|\varepsilon|^{2N-2}\varepsilon.\nonumber
\end{eqnarray}
where $\omega$ is the angular frequency, $n$ is the refractive index, $c$ is the light velocity, $m_\textrm{e}$ and $e$ are the electron mass and charge respectively, $n_2$ is the nonlinear refractive index, $\rho_{\textrm{at}}$ is the atoms density, $N$ is the ionization nonlinearity, $\sigma_N$ is the ionization cross-section, and $\nu_\textrm{e}$ is the effective collision frequency. We chose $N$=7.5 since it is the widely admitted effective nonlinearity driving the ionization rate in argon. Finally, we considered the same gas pressure (0.4 bar) than the one used during our experiments.

In the two-beam case, two gaussian pulses propagating slightly off-axis ($\theta=\pm$3$^\circ$) with respect to the $z$ axis and crossing along the $x$ dimension have been considered. The initial electric field then writes as

\begin{equation}
\varepsilon(x,y,t,z=0)=F(t)G(y)(\varepsilon_+H_+(x)+\varepsilon_-H_-(x))
\end{equation}
with
\begin{eqnarray}
G(y)&=&e^{-\frac{y^2}{\sigma_y^2}}e^{-ik_0\left(R_z-\sqrt{R_z^2-y^2}\right)},\\ \nonumber
H_\pm(x)&=&e^{-\frac{(x\pm x_0)^2}{\sigma_x^2}}e^{-ik_0\left(R_z-\sqrt{R_z^2-(x\pm x_0)^2}\right)}e^{\mp ik_0\sin\theta(x\pm x_0)},\\ \nonumber
F(t)&=&e^{-\frac{t^2}{\sigma_t^2}}, \sigma_x=\sigma_y=\sigma_0\sqrt{1+\frac{z_0^2}{z_r^2}},\\ \nonumber
R_z&=&z_0\left(1+\frac{z_r^2}{z_0^2}\right), z_r=\frac{\pi\sigma_0^2}{\lambda_0}, x_0=-z_0\tan(\theta),\\ \nonumber
\varepsilon_\pm &=&\sqrt{\frac{2}{\pi}\frac{P_\pm}{\sigma_x\sigma_y}}, P_\pm =\sqrt{\frac{2}{\pi}}\frac{E_\pm}{\sigma_t}
\end{eqnarray}
and $\sigma_0$=22 $\mu$m, $\theta$=3$^\circ$, $\lambda_0$=800 nm, $z_0$=-4 mm, and $\sigma_t$=85 fs, and $E_{\pm}$=25 $\mu$J.

\subsubsection{Discussions}
\begin{figure}[htbp!]
\begin{center}
      \includegraphics[keepaspectratio,width=8cm]{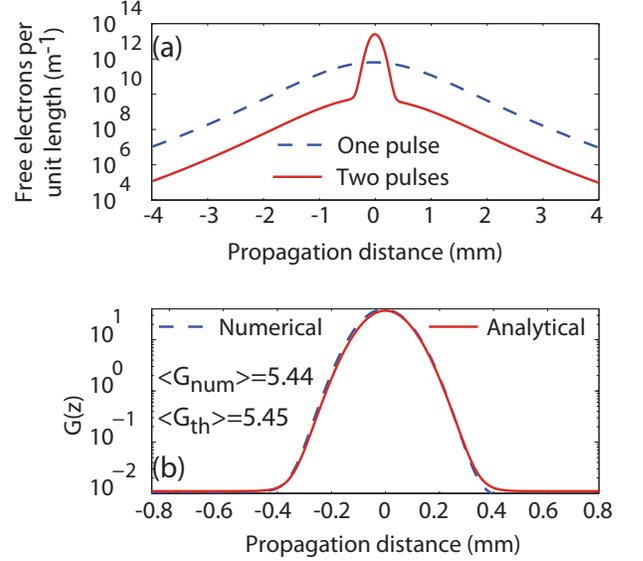}
\end{center}
   \caption{(Color online) (a) Amount of free electrons per unit length as a function of the propagation distance calculated according to the full 3D propagation equation in the one beam (dashed blue) and two-beam configurations (solid red). (b) Comparison between analytical and numerical gain as a function of the propagation distance.}
\label{Fig:3}
\end{figure}

Figure \ref{Fig:3}(a) displays the amount of free electrons per length unit generated in the single pulse case (blue dashed line) and in the two-pulse configuration (solid red line) calculated with the help of the numerical simulations. As expected, the production of free electrons is enhanced as soon as the pulses overlap. The maximal gain, which is obtained when the two pulses perfectly overlap ($z$=0), is about 37 in perfect agreement with Eq. \ref{Eq:Gplane}, which indicates that the plane wave approximation is valid in this particular situation. A remarkable agreement between analytical and numerical results is found when comparing the evolution of the free electrons gains all along the propagation as shown in Figure \ref{Fig:3}(b). The gain of free electrons averaged over the whole propagation obtained analytically (Eq. \ref{Eq:GaussianWave}) and numerically only differs by 0.2 $\%$ (5.45 and 5.44, respectively). It ensures that nonlinear propagation effects do not significantly alter the linear propagation, at least in the present initial energy and pressure conditions, and consequently validates the analytical results. The proposed method is therefore well suited for  fast and accurate determinations of unknown nonlinearity for experimental conditions where nonlinear propagation effects remains negligible.
\section{Experimental results}
\subsection{Experimental setup}
\begin{figure}[htbp!]
\begin{center}
      \includegraphics[keepaspectratio,width=9cm]{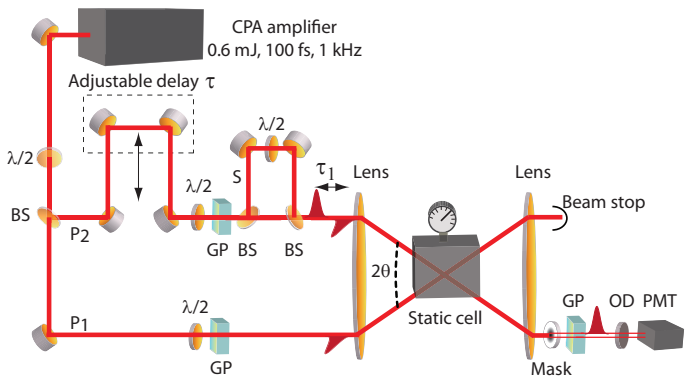}
\end{center}
   \caption{(Color online) Top view of the experimental setup. BS: beam splitter. GP: Glan cube polarizer. OD: optical density. P$_1$ (P$_2$): path 1 (2). S: probe.}
\label{Fig:4}
\end{figure}

The setup is depicted in Fig. \ref{Fig:4}. The optical source is a 1 kHz amplified femtosecond laser delivering vertically polarized 600 $\mu $J, 100 fs pulses at 796.3 nm. The beam is split into two parts, the relative delay $\tau$ between the two paths (P$_1$ and P$_2$) being controlled by means of a motorized delay line. The energy in the two optical paths is adjusted by means of an half-wave plate and a polarizer. In the path P$_2$, a Mach-Zehnder interferometer is inserted to produce two co-propagating pulses. The first one is used as a pump during the two-pulse experiments and is blocked for single pump experiments while the second one, delayed by around $\tau_1$=500 fs, is used to probe the plasma density generated by the single or the two-pump beam, depending on the experiment. The energy of the probe is about 4 $\%$ of the total energy in P$_2$. Moreover, its polarization is rotated by 90$^\circ$ with respect to the polarization of the pumps. The three pulses are then focused in a static cell. P$_1$ and P$_2$ are crossed with an angle 2$\theta\simeq$5.2$^\circ$. The waist of the pulses, $\sigma_\textrm{exp}=22\ \mu$m, were measured at weak intensity with the help of a camera. After the static cell, the probe is selected with the help of a polarizer. The plasma density is measured by the cross-defocusing technique as described in \cite{LoriotIonization,RenardDefocusing}. As shown in \cite{LoriotIonization,RenardDefocusing} in the case of parabolic pulses, the defocusing signal is proportional to $\Delta n^2$, \textit{i.e.} the peak to valley change of refractive index experienced by the probe beam. This is confirmed in the gaussian pulse case by our full 3D+1 numerical calculations simulating the experimental defocusing setup. In this experiment, the cross-defocusing signal is then proportional to the square of refractive index change resulting from the ionization mechanism and accordingly proportional to the square of the amount of free electrons generated during the experiment. It then allows a direct experimental measurement of the latter. Since the defocusing signal induced in the two-pulse scheme is about 30 times higher than the one induced during the single pulse experiment, a calibrated optical density is added before the photomultiplier during the former in order to keep the detection dynamics constant and to avoid a saturation of the photomultiplier.

Note that, in the two-beam case, the defocusing signal in the probe propagation direction is sensitive to the plasma density averaged over an interference fringe (as calculated in the analytical section) while the modulation due to interferences is responsible for the redirection of a small part of the probe \cite{GratingDiffraction}.
\subsection{Results and discussion}

\begin{figure}[htbp!]
\begin{center}
      \includegraphics[keepaspectratio,width=8cm]{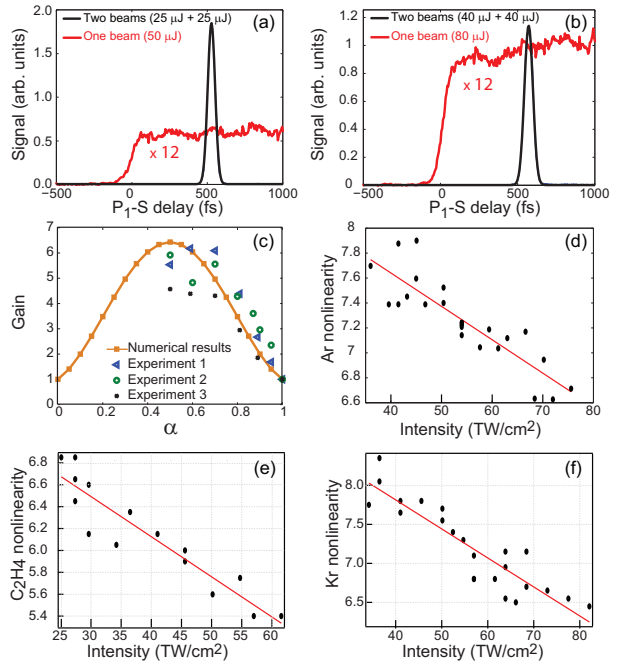}
\end{center}
   \caption{(Color online) Defocusing signal in argon at 50 $\mu$J (a) and at 80 $\mu$ J (b) as a function of the delay between pump and probe beams for the single and two-beam pump experiments. (c) Free electrons gain as a function of $\alpha$ in argon at 50 $\mu$J. Experiments 1, 2, and 3 correspond to three independent measurements sets. Measured nonlinearity as a function of laser peak intensity for (d) argon, (e) ethylene, and (f) krypton.}
\label{Fig:5}
\end{figure}

Figure \ref{Fig:5}(a), resp. (b), shows the defocusing signal in argon (0.4 bar) in both one (50 $\mu$J, resp. 80 $\mu$J, black line) and two-pulse  (25 $\mu$J+25 $\mu$J, resp. 40 $\mu$J+40 $\mu$J, red line) cases as a function the pump-probe delay. Note that it was checked with the help of 3D+1 numerical propagation simulations that intensity clamping does not occur at this pressure up to 90 TW/cm$^2$. As far as the single pulse case is concerned, the defocusing signal increases as soon as the pump and probe pulses temporally overlap and then remains roughly constant for the rest of the scan, as expected from plasma-induced defocusing (both electrons recombination and diffusion are negligible over the considered temporal window). In the two-pulse experiment, the defocusing signal remains hardly distinguishable on the scale of the figure until the two pulses temporally overlap, resulting in a sharp increase of the signal. The gain on the production of free electrons is retrieved from the experimental curves by comparing the post-pulse defocusing signal induced by a single pump to the one induced when the two pumps perfectly temporally overlap. The result obtained at $E_\textrm{tot}=50\mu$J as a function of $\alpha$ is depicted in Fig. \ref{Fig:5}(c). A nonlinearity $N_{\textrm{eff}}$=7.4$\pm$ 0.05 is found when fitting the experimental data with the analytical model (Eq. \ref{Eq:GaussianWave}), in good agreement with previous works performed in the same intensity regime.

The above effective nonlinearity is  intensity dependent. When an electron oscillates in the field, it acquires a ponderomotive energy that prevents the electron to escape from the atomic potential, making the ionization harder. Furthermore, as the electric field increases tunnel ionization takes over resulting in a deviation from the multiphoton ionization regime. Both effects are expected to reduce the effective nonlinearity as the intensity increases. In order to study such a dependence, the global energy was changed from about 20 $\mu$J up to 80 $\mu$J. Figures \ref{Fig:5}(d-f) show the nonlinearity of the ionization process as a function of the global energy in argon, ethylene, and krypton. These gases have been chosen because their ionization potential $I_\textrm{p}$ differs by several photon energy units (see Table \ref{Tab:1}) (E$_\textrm{ph}\simeq1.56$ eV). For all gases, the effective nonlinearity decreases by more than one unit as the intensity increases from 20 to 80 TW/cm$^2$ and can be fitted by a linear function $N_\textrm{exp}(I)=-\frac{I}{I_0}+N_0$ with respect to intensity $I$. The coefficients of the fit are summarized in Table \ref{Tab:1}. Finally, one has to emphasize that the nonlinearity measurements are performed by two measurements only (one and two-pulse configuration) unlike conventional technics. This technique then provides a very convenient method to study highly nonlinear systems.

\begin{table}[htbp!]
  \begin{tabular}{|c|c|c|c|}
  \hline
  & C$_2$H$_4$ & Kr & Ar \\
  \hline
  I$_\textrm{p}$ (eV) & 10.51 & 14.00 & 15.76 \\
  \hline
  N$_\textrm{th} @796.3\ \textrm{nm} $ & 7 & 9/10$^{(*)}$ & 11 \\
  \hline
  N$_\textrm{exp}$ @ 50 \textrm{TW/cm}$^2$ & 5.7 & 7.5 & 7.4 \\
  \hline
  $I_0$ (TW/cm$^2$) & 27.3 & 26.9 & 37.3 \\
  \hline
  $N_0$ & 7.6 & 9.3 & 8.7\\
  \hline
  \end{tabular}
  \caption{\label{Tab:1} Comparison between the theoretical nonlinearity in the pure multiphoton regime and the nonlinearity experimentally measured. $^{(*)}$ In the case of Kr, the theoretical nonlinearity at 796.3 nm is 9 while it is 10 at 797 nm, both wavelengths being covered by the laser bandwidth.}
\end{table}

These results could be used to improve the numerical models of ionization used, for instance, in filamentation studies. Nevertheless, since the presented method only gives the nonlinearity of the process but is unable to determine the absolute ionization probability, it should be coupled to a self-referenced method. For instance, by comparing the ionization induced refractive index change with the one induced by molecular alignment \cite{LoriotIonization}, it could be possible to determine the absolute cross-section of the ionization process.
\section{Conclusion}
In conclusion, a simple and fast all-optical method for measuring nonlinearities and optimizing their concomitant process is presented. The principle of the present method is validated both analytically and numerically. It is applied to the nonlinear photoionization of several gases. Moreover, it is shown that interferences between two crossing pulses lead to a net gain that can reach several order of magnitude on the nonlinearity generation efficiency at constant energy level. Finally, one has to emphasize that larger enhancement can be achieved by using more that two pulses. For instance, the maximal gain expected in the three plane waves configuration is $G_{3\ \textrm{beams}}=3^{-N}\ {}^2\!F_1\left(1/2-N,-N,1,4\right)$, the energy needed to produce a nonlinear effect being reduced up to a factor three. This improvement would require a stabilized interferometer for maximizing multi-waves interferences.
\acknowledgments
This work was supported by the Conseil R\'egional de Bourgogne (FABER program), the CNRS, and the LABEX ACTION. P.B. thanks the CRI-CCUB for CPU loan on its multiprocessor server.\vfill
\bibliographystyle{unsrt}

\end{document}